\documentclass{PoS}
\usepackage{bm}
\usepackage{epstopdf}
\usepackage{color}
\usepackage{soul}
\setulcolor{blue}
\setstcolor{red}
\sethlcolor{yellow}

\title{Meson screening masses at finite temperature with Highly Improved Staggered Quarks}

\ShortTitle{Meson screening masses at finite temperature with Highly Improved Staggered Quarks}

\author{\speaker{Yu Maezawa}\footnote{Current address: Fakult\"{a}t f\"{u}r Physik, Universit\"{a}t Bielefeld, D-33615 Bielefeld, Germany}\\
       Physics Department, Brookhaven National Laboratory, Upton, NY 11973, USA\\
       E-mail: \email{ymaezawa@quark.phy.bnl.gov}}

\author{Alexei Bazavov\\
        Physics Department, Brookhaven National Laboratory, Upton, NY 11973, USA}

\author{Frithjof Karsch\\
        Physics Department, Brookhaven National Laboratory, Upton, NY 11973, USA\\
        Fakult\"{a}t f\"{u}r Physik, Universit\"{a}t Bielefeld, D-33615 Bielefeld, Germany}

\author{Peter Petreczky\\
        Physics Department, Brookhaven National Laboratory, Upton, NY 11973, USA}

\author{Swagato Mukherjee\\
        Physics Department, Brookhaven National Laboratory, Upton, NY 11973, USA}

\abstract{
We report on the first study of the screening properties of the mesonic
excitations with strange ($s$) and charm ($c$) quarks, specifically the ground states
of the pseudo-scalar and vector meson excitations for the $\bar{s}s$, $\bar{s}c$ and
$\bar{c}c$ flavor combinations, using the Highly Improved Staggered Quark action with
dynamical physical strange quark and nearly-physical up and down quarks. By comparing
with their respective vacuum meson masses and by investigating the influence of the
changing temporal boundary conditions of the valence quarks we study the thermal
modifications of these mesonic excitations. While the $\bar{s}s$ states show
significant modifications even below the chiral crossover temperature $T_c$, the
modifications of the open-charm and charmonium like states become visible only for
temperatures $T\gtrsim T_c$ and $T\gtrsim1.2T_c$, respectively.
}

\FullConference{31st International Symposium on Lattice Field Theory LATTICE 2013\\
		 July 29-August 3, 2013\\
		 Mainz, Germany}

\begin{document}

\section{Introduction}

In-medium properties of hadronic excitations are important for understanding the
 nature of the Quark Gluon Plasma (QGP) created in the relativistic heavy-ion
 collision experiments at RHIC and LHC \cite{Satz:2012zza}. 
Inside QGP the interaction between a heavy quark-antiquark pair gets weakened due to the screening
 effects of the intervening deconfined colored medium. 
This telltale signature of the presence of a color deconfined medium is expected to be manifested through 
 melting of heavy quarkonium states, bound states of a quark-antiquark pair. 
In particular, melting of the charmonium states has been proposed as
 a signal for the creation of the QGP in the heavy-ion collision experiments \cite{Matsui:1986dk}.  
Thus it is important to reveal the fates of these meson states at high temperatures.

The lattice simulations enable non-perturbative calculations of
 physical quantities from first-principle QCD. 
The lattice QCD formalism is based on the Euclidian space-time and
 mesonic properties, in general, extracted from the corresponding current-current correlation 
  functions along the temporal direction. 
It is, however, difficult to study in-medium properties of mesons
 using their temporal correlation functions since at
  non-zero temperatures the physical distance along the temporal direction 
  is limited by the inverse temperature scale, $1/T$. 
To study in-medium properties of the quarkonia from the temporal correlators, 
 several approaches have been proposed, such as the maximum entropy method, 
 the variational technique \emph{etc.} \cite{Mocsy:2013syh}.

Spatial correlation functions of mesonic excitations are free from such a limitation.  
The spatial correlation functions, $C(z,T)$ enable us 
 to investigate thermal modification of the mesonic excitations at
  non-zero temperatures, as they too contain information about the temperature dependence 
   of the spectral functions, $\rho(\omega,T)$, of the mesonic excitations \cite{Karsch:2012na}
\begin{equation}
C(z,T) = \int_0^\infty \frac{2d\omega}{\omega} \int_{-\infty}^\infty dp_z e^{i p_z z}
\rho(\omega,p_z,T)
\;.
\end{equation}
At large distances the spatial correlation functions decay exponentially
characterized by the decay constant know as the \emph{screening mass} ($M$). At small
enough temperatures when there exists a well-defined mesonic bound state with
$\rho(\omega)\sim\delta(\omega^2-p_z^2-m_0^2)$, then $M$ becomes equal to the (pole)
mass $m_0$ of the meson. On the other hand, at high enough temperatures when the
mesonic excitation is completely melted and consists of a free quark-antiquark pair
then $M_{\rm free}(T)=2\sqrt{(\pi T)^2 + m_q^2}$, where $m_q$ is the bare mass of the
valence quark. Such behavior of the screening mass in the non-interacting limit is
the consequence of the fact that at non-zero temperatures while traversing across the
temporal boundary the free (anti)quark picks up at least a $\pi T$ contribution from
the lowest Matsubara frequency due the anti-periodic temporal boundary condition
(APB). Thus the transition between these two limiting values of the screening mass
can used as an indicator for the thermal modification and eventual dissolution of the
mesonic excitations.

Moreover, if a valence quark-antiquark pair exists as a mesonic bound state then
due to the bosonic nature of this excitation its screening mass is not expected to be
sensitive to the anti-periodic (fermionic) nature of the temporal boundary condition
of its constituent (anti)quark. In such a scenario, even if one imposes a periodic
temporal boundary condition (PB) for the quarks the screening mass of the mesonic
excitation is expected to remain largely unchanged compared to the usual APB case. On
the other hand, for a non-interacting quark-antiquark pair with PB the screening mass
will be $M_{\rm free}(T)=2m_q$, instead of the $M_{\rm free}(T)=2\sqrt{(\pi T)^2 + m_q^2}$
for the case of usual APB. Thus, by studying the influence of the changing temporal
boundary condition of the quarks on the screening mass of the quark-antiquark system
we can get additional information on the possible existence of mesonic bound states
inside the QGP \cite{Karsch:2012na}.

\section{Simulation details}

Lattice QCD studies of the screening masses have been performed within 
 the quenched approximation \cite{DeTar:1987xb,de Forcrand:2000jx}
 and also in full-QCD using the staggered-type quarks 
  \cite{Karsch:2012na,Boyd:1994np,Pushkina:2004wa,Cheng:2010fe}
 as well as using Wilson-type quarks \cite{Iida:2010jz}.  
In this work we report, for the first time, studies of the meson screening masses
 using the Highly Improved Staggered Quark (HISQ) action 
  with dynamical physical strange quark and nearly-physical up and down quarks. 
At a given lattice spacing the HISQ action is known to have
 taste symmetry violations that are smaller than those observed
 with all other staggered-type actions currently are used in studies
 of lattice QCD thermodynamics \cite{Bazavov:2011nk}.
In this work we study the spatial correlation functions of the mesonic excitations with
strange ($s$) and charm ($c$) quarks, specifically the ground states of the
pseudo-scalar and vector meson excitations for the $\bar{s}s$, $\bar{s}c$ and
$\bar{c}c$ flavor combinations.

We calculate the meson correlation functions on gauge configurations generated
with 2+1 flavor QCD using the HISQ action \cite{Bazavov:2011nk}. The strange
quark mass $m_s$ is adjusted to its physical value and the light quark masses are
fixed at $m_l=m_s/20$, corresponding to $m_\pi \simeq 160$ MeV and $m_K \simeq 504$
MeV at zero temperature. Charm quarks are introduced only as valance quarks. The
valence charm quark mass is adjusted to reproduce a spin-averaged charmonium mass,
$\frac{1}{4} ( m_{\eta_c} + 3 m_{J/\psi})$, at zero temperature.  The majority of the
calculations are performed using $N_\tau = 12$ lattices with the lattice couplings of
$\beta=6.664-7.280$ which corresponds to temperatures $T=138-245$ MeV. For $N_\tau =
10, 8, 6, 4$ lattices we adopt the fixed-scale approach by performing calculations at
a fixed $\beta=7.280$ with temperatures ranging from $T=297-743$ MeV.  For all
calculations the aspect ratios of the lattices are chosen to be $N_s/N_\tau = 4$. For
the chiral crossover temperature we use the continuum extrapolated value of
$T_c=154\pm9$ MeV \cite{Bazavov:2011nk}.

In the staggered formulation a quark contains four valence tastes 
 and meson operators are defined as ${\cal M} = \bar{q} ( \Gamma^D \times \Gamma^F ) q$, 
 $\Gamma^D$ and $\Gamma^F$ being products of the Dirac Gamma matrices which generate spin and 
 taste structures, respectively \cite{Altmeyer:1992dd}.  
In this study, we focus only on local meson operators with $\Gamma^D = \Gamma^F = \Gamma$.  
By using staggered quark fields $\chi({\bm x})$ at ${\bm x}=(x,y,z,\tau)$ the meson operators
 can be written as ${\cal M}({\bm x})=\tilde{\phi}({\bm x}) \bar{\chi}({\bm x})
 \chi({\bm x})$, where $\tilde{\phi}({\bm x})$ is a phase factor depending on the
 choice of $\Gamma$. 
We calculate only the quark-line connected part of the meson correlators 
 since the effect due to the disconnected part is expected to be small. 
The spatial correlation function along the $z$ direction, projected to zero transverse momentum,
is obtained as
\begin{eqnarray}
C(z) = \sum_{x,y,\tau} \phi({\bm x}) \left\langle \left( M^{-1}_{{\bm 0} {\bm x}} \right)^\dag M^{-1}_{{\bm 0} {\bm x}} \right \rangle ,
\end{eqnarray}
where $M^{-1}_{{\bm 0} {\bm x}}$ is the staggered fermion propagator from ${\bm 0}$
 to ${\bm x}$ and $\phi({\bm x})$ is the phase factor defined as $\phi({\bm x}) = -
 (-1)^{x+y+t}\tilde{\phi}({\bm x})$.  
Since a staggered meson correlator contains two different mesons with opposite parities, 
 at large distances the lattice correlator can be parametrized as 
\begin{eqnarray}
C(z) = A_{\rm NO} \cosh \left[ M_- \left(z-\frac{N_z}{2} \right) \right] - (-1)^z A_{\rm O} \cosh \left[ M_+ \left(z-\frac{N_z}{2} \right) \right],
\label{eq:fit}
\end{eqnarray}
where the first (second) term in the right-hand-side characterizes a non-oscillating
 (oscillating) contribution governed by a negative (positive) parity state.  
Since at non-zero temperatures the rotational symmetry of the spatial correlations functions 
 reduces to $O(2) \times Z(2)$, the vector meson states must be
 decomposed into transverse and longitudinal components. 
Here we only focus on the meson screening masses in the pseudo-scalar (PS$-$) and 
 vector-transverse (V$-$) channels with the negative parity, defined by
 $(\Gamma, \phi) = (\gamma_5, 1)$ and $(\gamma_{1,2}, (-1)^x \ {\rm and} \ (-1)^y)$, respectively. 
At zero temperature, these correspond to (PS$-$, V$-$) $= 
 (\eta_{\bar{s}s},\ \phi), \ (D_s, \ D_s^\ast), \ (\eta_c , \ J/\psi)$ for the
 $\bar{s}s$, $\bar{s}c$ and $\bar{c}c$, respectively, 
  where $\eta_{\bar{s}s}$ is an unphysical state with the mass 
   $M_{\eta_{\bar{s}s}} = \sqrt{2M_K^2 - M_\pi^2}$. 
We investigate thermal properties of these states by studying temperature dependence
 of the corresponding screening masses using both periodic and anti-periodic
 temporal boundary conditions for the valence quarks.

\begin{figure*}[t]
\begin{center}
\begin{tabular}{cc}
                   \includegraphics[width=73mm]{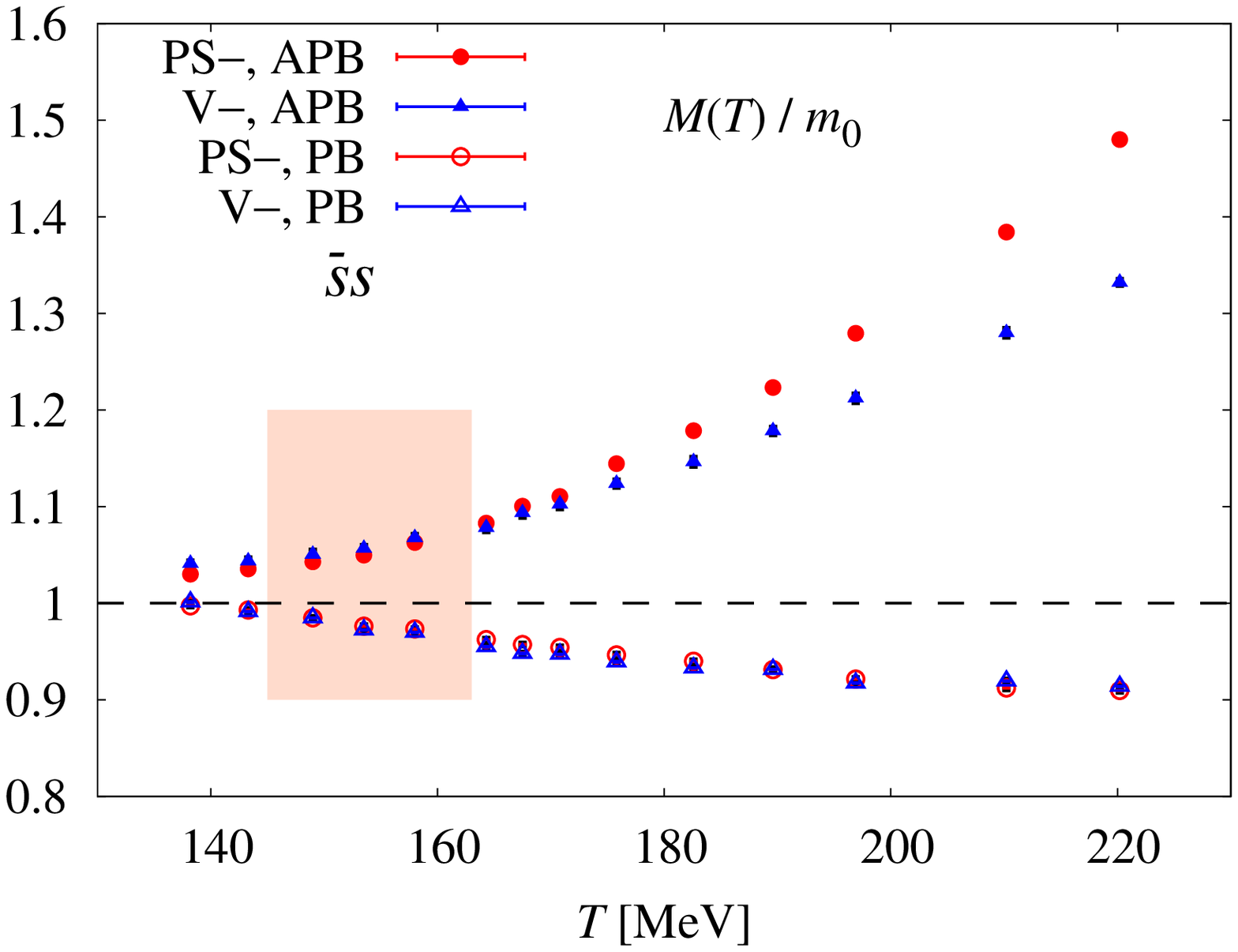} &
                   \includegraphics[width=73mm]{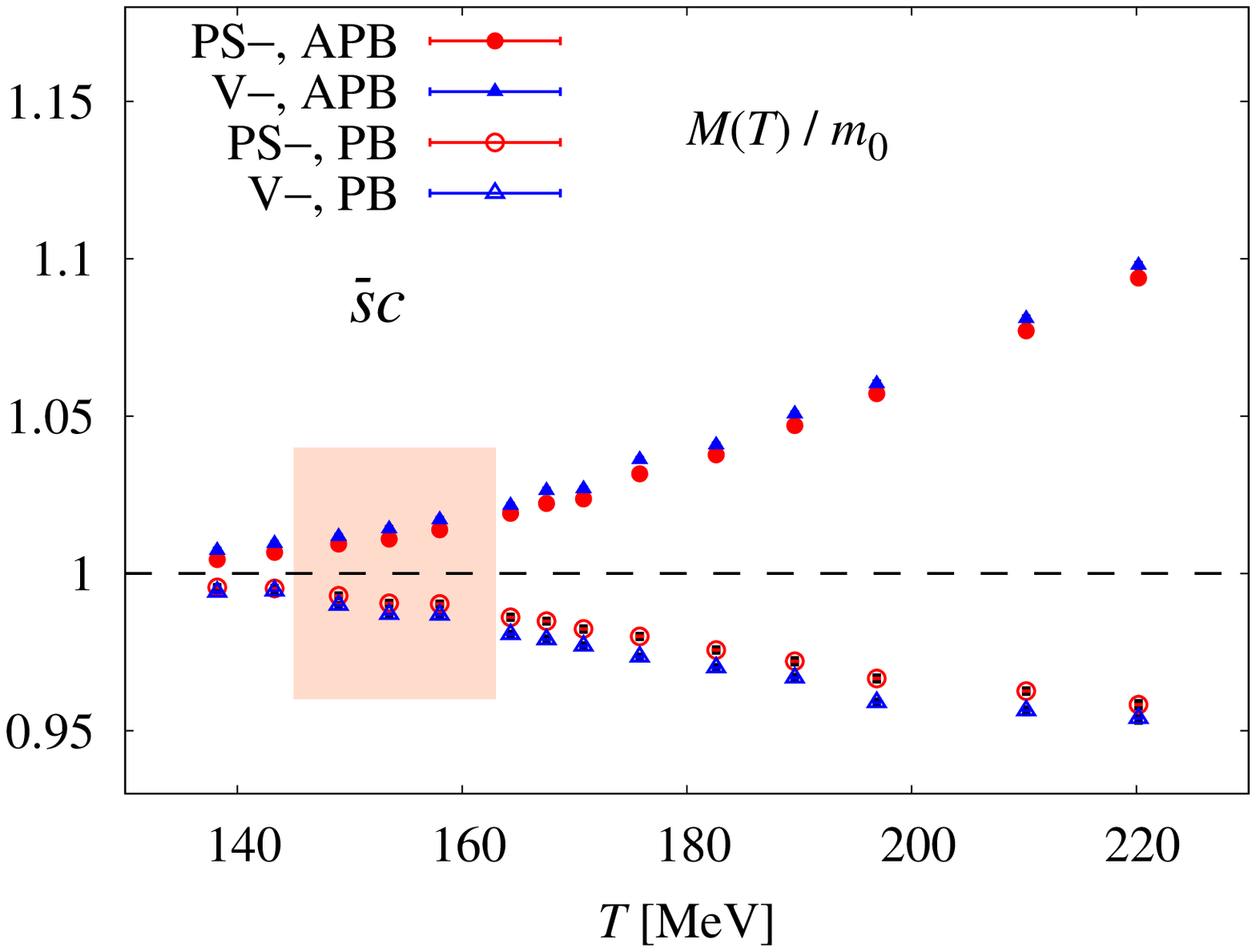} \\
\multicolumn{2}{c}{\includegraphics[width=73mm]{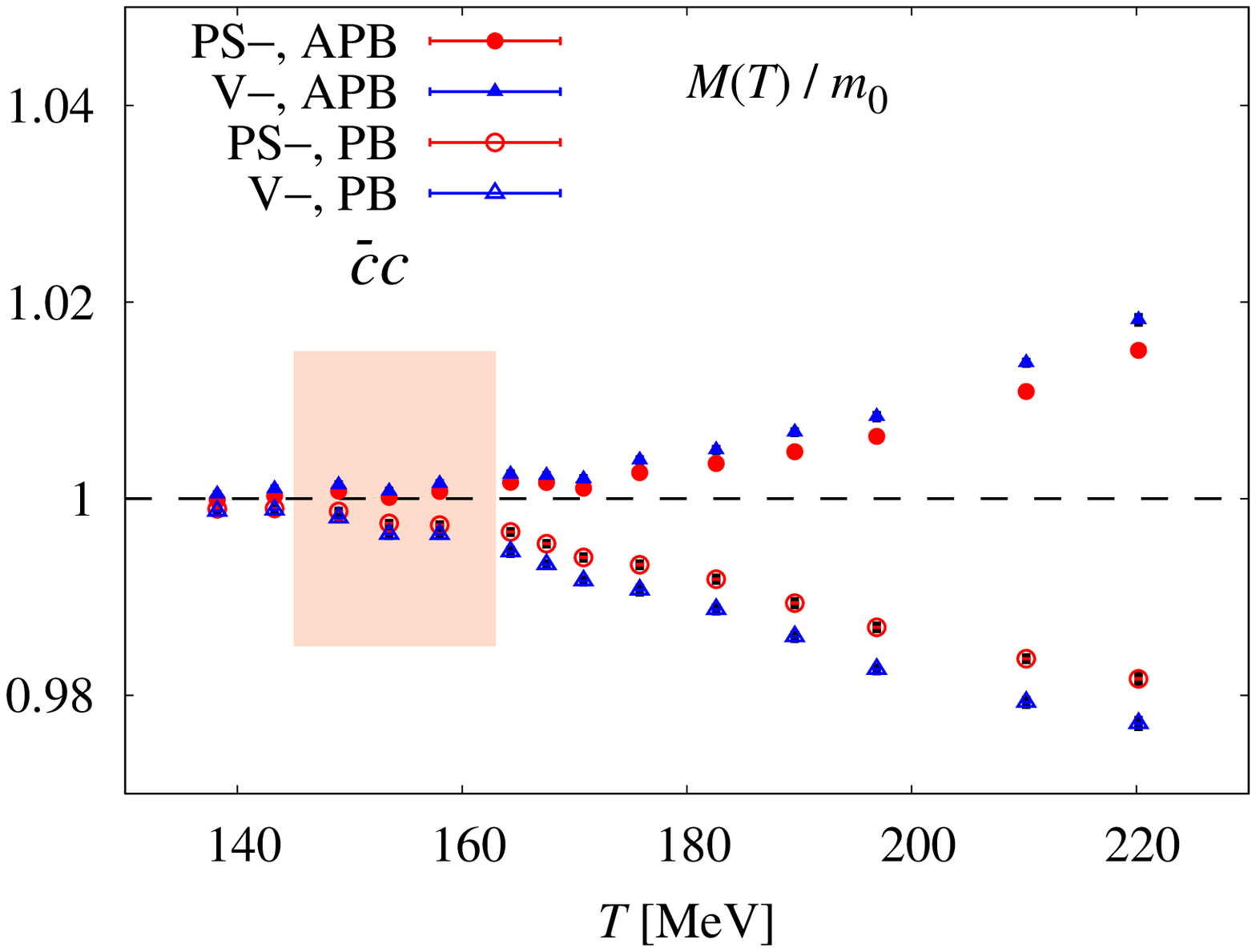}}
\end{tabular}
\caption{Ratios of the screening masses to the corresponding zero temperature masses
 for the $\bar{s}s$ (top-left), $\bar{s}c$ (top-right) and $\bar{c}c$ (bottom) sectors.} 
\label{fig:sm}
\end{center}
\end{figure*}

\begin{figure*}[t]
\begin{center}
\begin{tabular}{cc}
                   \includegraphics[width=73mm]{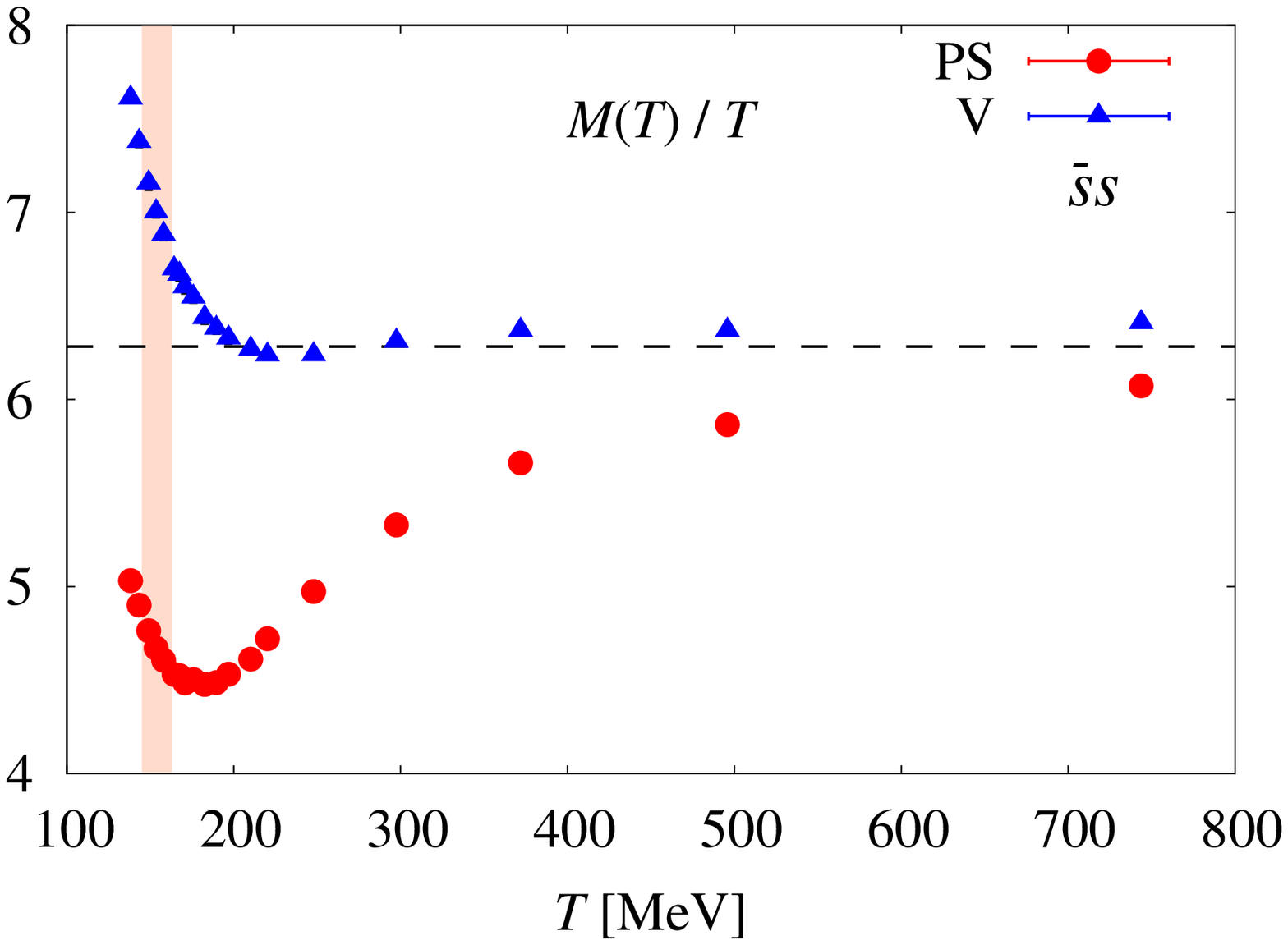} &
                   \includegraphics[width=73mm]{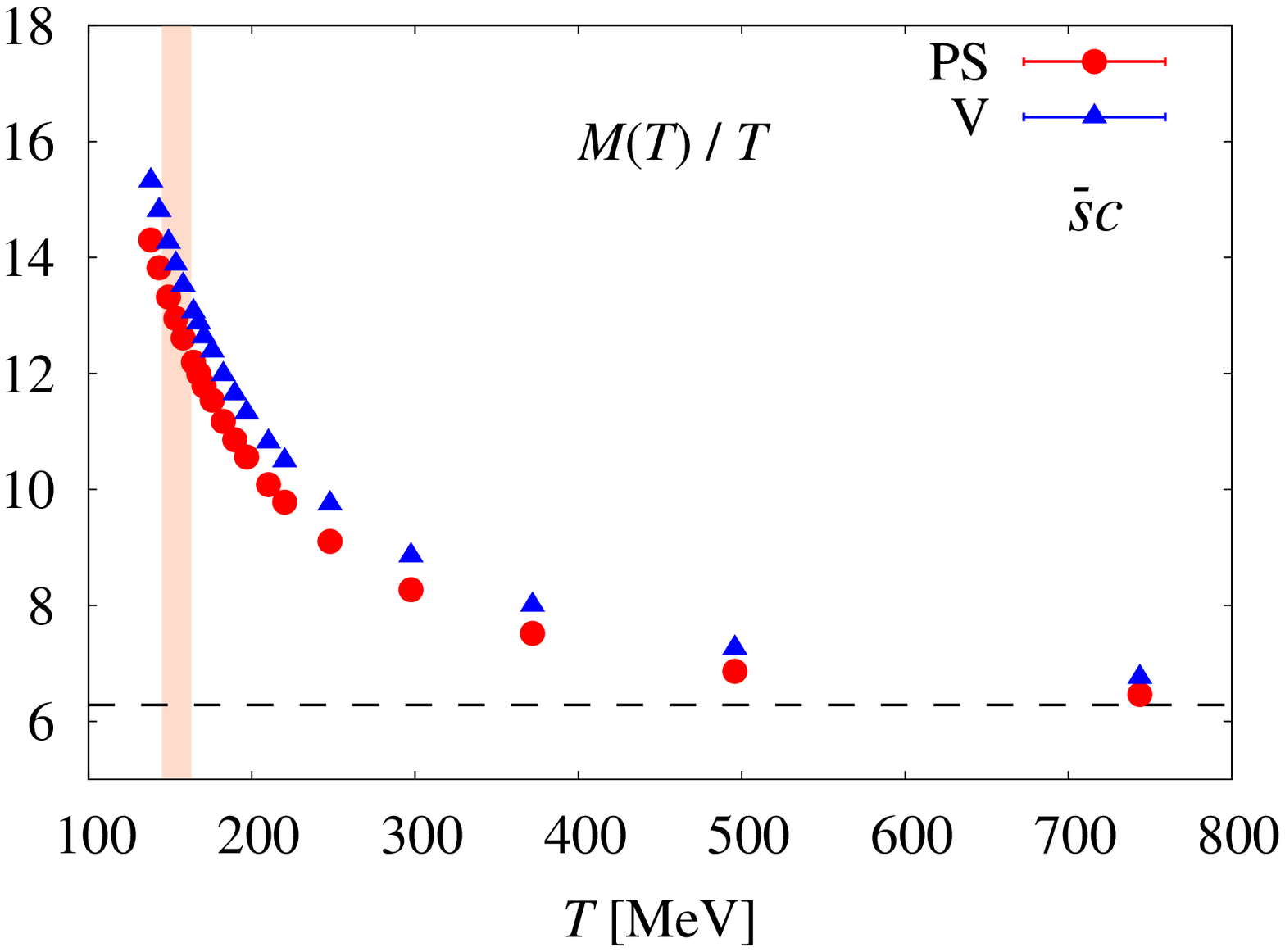} \\
\multicolumn{2}{c}{\includegraphics[width=73mm]{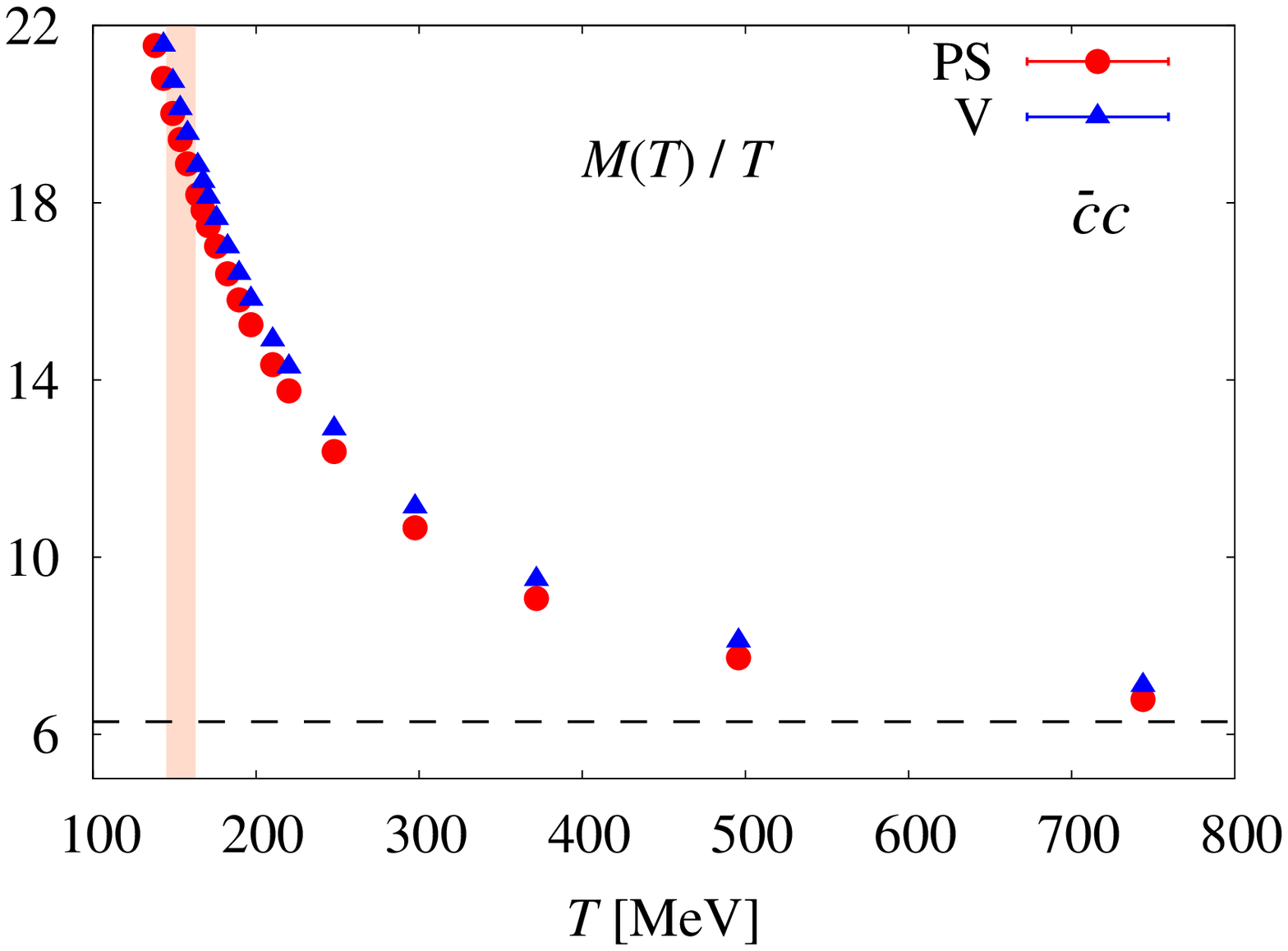}}
\end{tabular}
\caption{High temperature behavior of the screening masses, in units of temperature,
 for the $\bar{s}s$ (top-left), $\bar{s}c$ (top-right) and $\bar{c}c$ (bottom) sectors.}
\label{fig:small}
\end{center}
\end{figure*}

\section{Results}

We have calculated the spatial meson correlators for the $\bar{s}s$, $\bar{s}c$ and $\bar{c}c$ 
 flavor combinations and extracted the screening masses of the ground states in
 pseudo-scalar (PS$-$) and vector (V$-$) channels by fitting the correlators at large
 distances with Eq.~(\ref{eq:fit}) using Bayesian fits\footnote{We
 impose $A_{\rm O}=0$ on the PS$-$ correlators in $\bar{s}s$ and $\bar{c}c$} \cite{Lepage:2001ym}. 
Statistical errors have been estimated using bootstrap analysis. 

Figure \ref{fig:sm} shows the temperature dependence of the screening masses
 normalized by the corresponding meson (pole) masses at zero temperature,
 i.e. $M(T)/m_0$, for the $\bar{s}s$ (top-left), $\bar{s}c$ (top-right) and
 $\bar{c}c$ (bottom) sectors. 
Circles (triangles) indicate results for the PS$-$ and V$-$ channels, 
 and the filled (open) symbols indicate results obtained with APB (PB) temporal boundary
 conditions for the valence quarks.  The band denotes the chiral crossover temperature.  
The screening masses agree with the corresponding $T=0$ meson masses at low temperatures. 
Deviations between APB and PB conditions gradually set in as the temperature increases.  
At high temperatures, the screening masses with APB (PB) condition increase (decrease) 
 which is consistent with the behavior in the free limit, 
 i.e. $M_{\rm free} (T) = 2 \sqrt{(\pi T)^2 + m_q^2}$ for APB and $2m_q$ for PB, respectively. 
While the screening masses in the $\bar{s}s$ sector (top-left) 
 significantly differ from the corresponding $T=0$ meson masses even for $T<T_c$ 
 similar deviations for the $\bar{s}c$ (top-right) and $\bar{c}c$ (bottom) sectors start to
  appear around $T\gtrsim T_c$ and $T\gtrsim1.2T_c$, respectively.  
This may imply that the $\eta_{\bar{s}s}$ mesons receive significant thermal modifications
 even below $T_c$ but the $\eta_c$ and $J/\psi$ remain largely unaffected by the thermal medium 
 up to $T\lesssim1.2T_c$.

Figure \ref{fig:small} shows high temperature behaviors of the screening masses, 
 normalized by temperature $M(T)/T$, with APB condition
 for the $\bar{s}s$ (top-left), $\bar{s}c$ (top-right) and $\bar{c}c$ (bottom) sectors.
The dashed lines denote the value $2\pi$ to which the screening masses converge in
 the high temperature limit of the screening masses for massless non-interacting quarks. 
The $\bar{s}c$ and $\bar{c}c$ screening masses monotonically decrease with temperature 
 and tend to converge to the massless non-interacting limit of $2\pi$
 for $T\gtrsim 700$ MeV. 
On the other hand, the $\bar{s}s$ sector shows more interesting features: 
 the screening masses decrease at low temperature, start to increase again for $T\gtrsim T_c$
 and then tend to converge towards $2\pi T$ for $T\gtrsim700$ MeV with the  PS$-$
 (V$-$) channel approaching from below (above).

\section{Summary}

In this work we have investigated the screening masses of the ground state
pseudo-scalar and vector mesonic excitations for the $\bar{s}s$, $\bar{s}c$ and
$\bar{c}c$ flavor combinations. Calculations have been performed using 2+1 flavors of
dynamical HISQ along with valence charm quarks. We have investigated the signatures
of thermal modifications of these states by comparing their screening masses with the
corresponding zero-temperature meson masses. We have also studied the influence of
the changing temporal boundary conditions of the valence quarks on the screening
masses of these states. Combining these studies we have found clear signatures of
thermal modifications for the $\bar{s}s$ states even for $T<T_c$, $T_c=154$ MeV being
the chiral crossover temperature. For the $\bar{s}c$ and $\bar{c}c$ states
distinctive signatures of thermal modifications appear only for $T\gtrsim T_c$ and
$T\gtrsim1.2T_c$, respectively. On the other hand, for all sectors the screening
masses approach the non-interacting massless quark gas values of $2\pi T$ only at
very high temperatures of $T\gtrsim700$ MeV.

\section*{Acknowledgement}

Numerical calculations were carried out on the USQCD Clusters at
 the Jefferson Laboratory, USA and the NYBlue at the Brookhaven National
Laboratory, USA. 
This work was partly supported by through the Contract No. DE-AC02-98CH10886 with the U.S. Department of Energy.
Partial support for this work was also provided through Scientific
Discovery through Advanced Computing (SciDAC) program funded by U.S.
Department of Energy, Office of Science, Advanced Scientific Computing
Research (and Basic Energy Sciences/Biological and Environmental
Research/High Energy Physics/Fusion Energy Sciences/Nuclear Physics).

\end{document}